\documentclass[manuscript]{acmart}

\AtBeginDocument{%
  }

\setcopyright{acmlicensed}
\copyrightyear{2027}
\acmYear{2027}
\acmDOI{XXXXXXX.XXXXXXX}
\acmConference[Conference 'XX]{the 20XX Conference}{May 10--14,
  2027}{Pittsburgh, PA, USA}
 
\acmISBN{978-1-4503-XXXX-X/2018/06}

\usepackage{stfloats}
\usepackage{booktabs}
\usepackage{multirow}
\usepackage{array} 
\usepackage[table]{xcolor}

\usepackage{algorithm}
\usepackage{algpseudocode}

\usepackage{enumitem}
\setlist[itemize]{leftmargin=*}

\begin{document}

\newcommand{\sys}[0]{\textsc{ArguScale}}

\title[\sys{}]{\sys{}: Evidence-Grounded Human-AI Sensemaking Across Concurrent Live Conversations}

\author{Panayu Keelawat}
\affiliation{%
  \institution{Virginia Tech}
  \city{Blacksburg}
  \state{Virginia}
  \country{USA}}
\email{panayu@vt.edu}

\author{Caleb Wohn}
\affiliation{%
  \institution{Virginia Tech}
  \city{Blacksburg}
  \state{Virginia}
  \country{USA}}
\email{cwohn@vt.edu}

\author{Shiva Pochampally}
\affiliation{%
  \institution{Virginia Tech}
  \city{Blacksburg}
  \state{Virginia}
  \country{USA}}
\email{shivapochampally@vt.edu}

\author{Rezky Dwisantika}
\affiliation{%
  \institution{University of Minnesota - Twin Cities}
  \city{Minneapolis}
  \state{Minnesota}
  \country{USA}}
\email{dwisa001@umn.edu}

\author{Eugenia H. Rho}
\affiliation{%
  \institution{Emory University}
  \city{Atlanta}
  \state{Georgia}
  \country{USA}}
\email{eugenia.h.rho@emory.edu}

\author{Yan Chen}
\affiliation{%
  \institution{Virginia Tech}
  \city{Blacksburg}
  \state{Virginia}
  \country{USA}}
\email{ych@vt.edu}

\renewcommand{\shortauthors}{Keelawat et al.}


\begin{abstract}
Monitoring concurrent classroom collaborative reasoning discussions is challenging because instructors have to track collaboration-relevant behaviors across multiple ongoing conversations at once. Existing Human-AI conversational sensemaking systems can scale monitoring upward through summaries and visual abstractions, but these representations may distance users from underlying conversational evidence, e.g., relevant audio and transcript segments, needed to forage downward. Through our formative study, we developed \sys{}, an evidence-grounded Human-AI sensemaking system that organizes interactional signals into temporally aligned representations linked to transcript and audio evidence. A temporal signal-density view provides temporal context and supports rapid cross-group comparison, while selective evidence access allows instructors to inspect audio and transcript. We evaluated \sys{} in a mixed-design study with 20 participants. Compared to other conditions, \sys{} improved participants' ability to efficiently and accurately identify collaboration-relevant cues. We further identify strategies participants used to forage for and make sense of conversational evidence under live time constraints.
\end{abstract}



\begin{teaserfigure}
  \centering
  \includegraphics[width=1\textwidth]{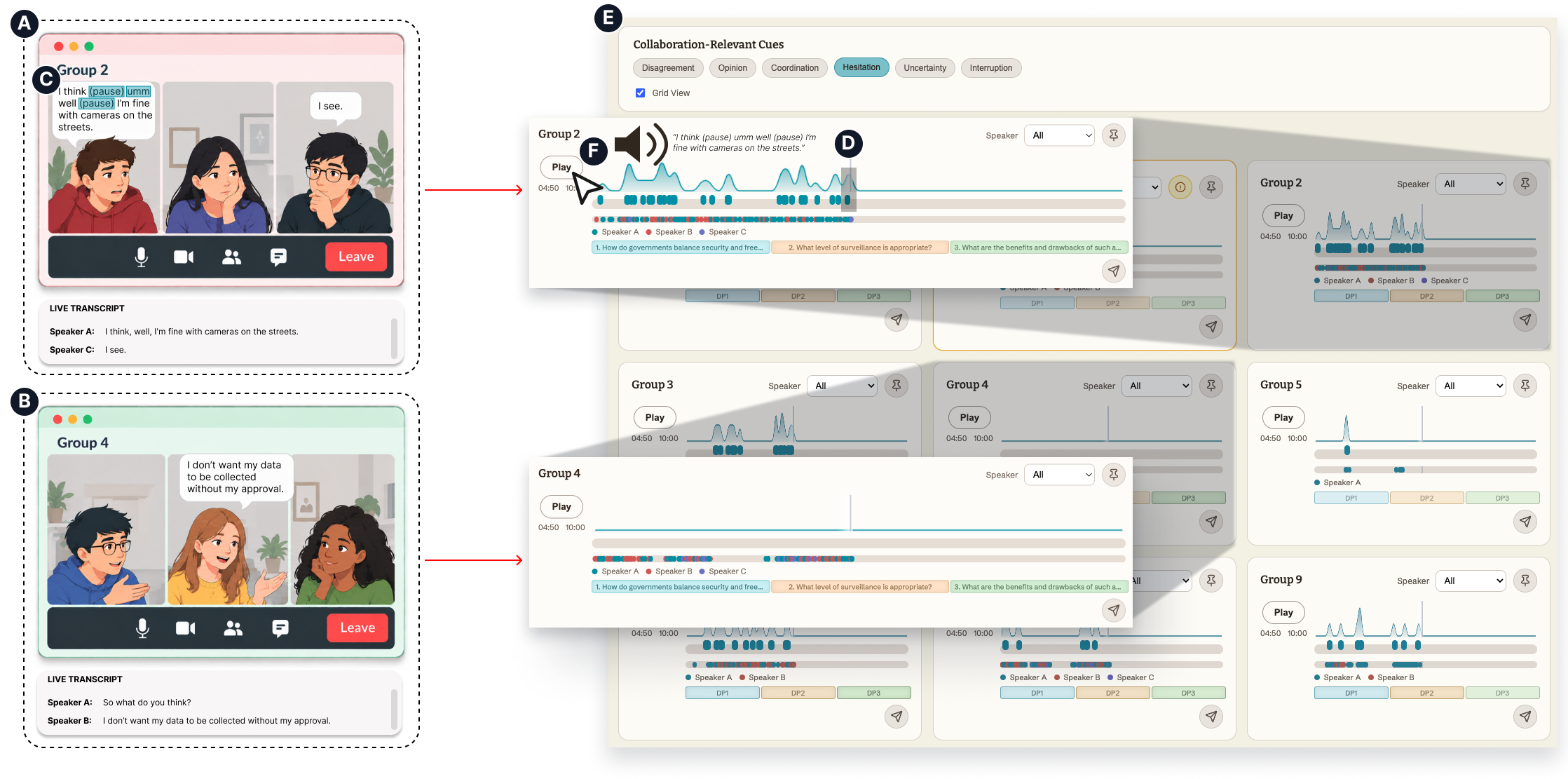}
  \caption{\sys{} supports evidence-grounded sensemaking of concurrent classroom collaborative reasoning discussions by processing and visualizing collaboration-relevant cues. In this example, students are discussing the topic of ``Public Surveillance and Privacy'' during an online class.
  (A) Group 2 contains a hesitant contribution, illustrated by pauses and fillers in the discussion example, while 
  (B) Group 4 shows a contribution without hesitation. 
  (C) The hesitant speech in Group 2 contains acoustic evidence, such as pauses and fillers, that is sometimes not visible in a typical transcript.
  (D) In \sys{}, this evidence is reflected along the group timelines. 
  (E) Instructors can select which collaboration-relevant cues to monitor across groups while viewing all groups; here, \textit{Hesitation} is selected. 
  (F) Instructors can further inspect a detected moment by reviewing its transcript and listening to the corresponding audio evidence.}
  \label{fig:teaser}
\end{teaserfigure}
\maketitle

\section{Introduction}

Collaborative reasoning (CR) is an important pedagogical approach that enables students to think critically about open-ended topics while developing the interpersonal skills needed to engage respectfully with diverse perspectives \cite{clark2003collaborative, kumpulainen2003interpersonal}. This is especially critical in computing, where technologies increasingly shape society and require careful consideration of what should be built, what should not, and their broader societal impacts \cite{embededEthics, UsingPackback}. For example, Fig.~\ref{fig:teaser} shows students discussing ``Public Surveillance and Privacy,'' where each group collectively considers how to balance government control with individual privacy. As a result, ethics curricula are becoming a core requirement in CS education \cite{ABET, TeachingEthicsInCS}. While CR can take place through class-wide discussions, small-group discussions are commonly used to promote more balanced participation in large classrooms \cite{techEthicsSyllabi}. In a typical workflow, instructors introduce a topic, often accompanied by readings, provide guiding discussion points, divide students into small groups, monitor group discussions, and conclude with a class-wide debrief. However, this structure creates a ``one-to-many'' challenge in which instructors have to oversee multiple parallel discussions. As a result, it becomes difficult for instructors to make sense of how each group is progressing through the activity, especially in online breakout-room settings where visibility into group discussions is even more limited.

Recent Human-AI sensemaking systems have addressed this scalability challenge by transforming large amounts of conversational data into higher-level representations that can be inspected more efficiently \cite{GroupTrackVis, Groupnamics, Dynamite}. However, sensemaking is a bidirectional process \cite{pirolli2005sensemaking}. While AI can support the upward process of abstracting raw conversations into summaries, structures, and visual representations, users also need to move downward from these abstractions to examine the evidence underlying an interpretation \cite{PaperTrail}. Prior systems have supported such evidence grounding by linking higher-level representations back to conversational sources such as transcripts \cite{Orality, MeetMap, TalkTraces}, while systems such as CPVis demonstrate the value of richer multimodal evidence for understanding collaborative behavior in post-hoc performance analysis \cite{CPVis}. Yet grounding interaction evidence across concurrent live conversations introduces a tension: users have to inspect evidence quickly enough while keeping pace with other conversations that continue to evolve in parallel. While systems such as Dynamite \cite{Dynamite} and MobileGroupVis \cite{MobileGroupVis} visualize artifacts derived from live group discussions, their underlying evidence is based on transcribed text, which can abstract away behavioral information conveyed through acoustic cues. Such cues may be important for instructors to understand how CR unfolds and may require listening to the underlying conversational audio evidence to interpret collaboration-relevant cues, such as hesitation or uncertainty, that cannot be fully represented through text alone.

To better understand the design challenges, we first interviewed two ethics instructors at our institution about the kinds of information they would want to hear from student groups during live discussions. We then observed their classroom practices and, with student consent, collected conversational data from three course sessions involving 85, 35, and 34 students. Based on these observations and recordings, we developed an initial aggregated prototype that spatially clustered related discussion content and associated detected interaction signals with corresponding audio snippets as evidence. We then asked three teaching assistants (TAs) to use the prototype as a sensemaking tool for monitoring simulated live small-group discussions and interviewed them afterward about their experience. Despite having direct access, participants rarely interacted with the evidence and instead relied on summaries maintain awareness across groups. We found that participants were not able to determine whether a detected interaction was an isolated occurrence or part of a recurring pattern because the prototype lacked sufficient temporal context. The findings highlighted the difficulty of making rich conversational evidence practically accessible under live time constraints and informed the design goals and subsequent iterations of \sys{}.

We therefore developed \sys{}, an evidence-grounded Human-AI sensemaking system that supports instructors in monitoring concurrent live conversations while retaining access to the underlying conversational evidence. \sys{} detects collaboration-relevant cues from ongoing discussions and places them along each group's conversational timeline. To support rapid comparison across groups, the system further summarizes the temporal concentration of these cues using a temporal signal density plot, where peaks indicate periods with higher concentrations of detected cues. \sys{} uses these AI-generated representations as intermediate representations between transcripts and raw audio, providing a bridge to the underlying audio evidence. Instructors can inspect individual cues, review the corresponding transcript, and selectively listen to the associated audio segment when additional acoustic context is needed. Together, these features are designed to reduce the cost of locating and inspecting relevant evidence while allowing instructors to maintain awareness of multiple conversations that continue to unfold in parallel.

We evaluated \sys{} in a mixed-design in-lab study with 20 participants comprising instructors and TAs. The study compared a transcript baseline, an ablated version of \sys{} without interactional cue visualizations, and the full \sys{} system, with all conditions retaining access to source audio. We examined two research questions: \textbf{RQ1:} \textit{How does \sys{} affect users' ability to locate and inspect relevant conversational evidence during live sensemaking across concurrent discussions?} and \textbf{RQ2:} \textit{What evidence-foraging and grounding strategies do users adopt while making sense of multiple concurrently evolving conversations with \sys{}?} Quantitatively, the full \sys{} system improved the speed and accuracy with which participants identified CR-related evidence. Qualitatively, participants used the representations to manage the temporal cost of evidence inspection through three recurring strategies: \textit{deferred grounding}, preserving a moment for later investigation; \textit{evidence revisitation}, returning to prior moments as their understanding evolved; and \textit{progressive grounding}, inspecting lightweight evidence such as transcripts before committing additional attention to audio.

Our contributions are three-fold:
\begin{itemize}
    \item Through a formative study with instructors and TAs, we identify key bottlenecks in evidence-grounded sensemaking across concurrent live conversations and derive design requirements for supporting rapid and selective access to conversational evidence under live time constraints.
    \item We introduce \sys{}, an evidence-grounded Human-AI sensemaking system that uses temporally aligned AI-generated intermediate representations to make conversational moments persistent, selectively inspectable, and directly groundable in transcript and audio evidence.
    \item We provide empirical findings from a mixed-design study showing improved efficiency and accuracy in grounding interactional interpretations, and identify deferred grounding, evidence revisitation, and progressive grounding as strategies for navigating conversational evidence during concurrent live discussions.
\end{itemize}

\section{Related Work}

\subsection{Collaborative Reasoning in Classrooms}

CR brings together both collaboration and reasoning, requiring students not only to articulate their own thinking but also to build understanding with others \cite{clark2003collaborative}. CR is practiced across educational settings, from K--12 classrooms to college-level instruction \cite{anEthicsPrimer}. During these discussions, students contribute claims, evidence, and explanations while engaging with peers to surface and examine different perspectives \cite{DiscussionTracker,groupCommAnalysis}. Instructors may therefore have expectations about the discussion content that groups should cover \cite{embededEthics}. However, the interaction dynamics within each group are also important because they can shape how students learn from one another \cite{kumpulainen2003interpersonal,cs1GroupingSIGCSE}. These dynamics can provide instructors with useful information for inferring how students are engaging with the activity and where they may be experiencing difficulties \cite{brennan1995feeling,prosodyStorytelling}. While CR can be conducted as a class-wide, open-floor discussion, this format may limit participation, leaving much of the class silent \cite{LLMClassDebateCHI,techEthicsSyllabi}. Small-group CR can create more opportunities for students to contribute, but it introduces a one-to-many challenge in which instructors have to monitor multiple discussions at once.

Our work investigates how to support instructors in making sense of these evolving conversations during live small-group CR activities. In particular, we examine how instructors can keep track of both what groups are discussing and how students are interacting with one another, so that monitoring reflects both the reasoning and collaboration aspects of CR.

\subsection{Conversation Sensemaking Tools}

A first group of systems supports sensemaking within a single conversation. TalkTraces visualizes conversational themes, Orality organizes spoken thoughts into a semantic canvas, and MeetMap constructs dialogue maps from ongoing meetings \cite{TalkTraces,Orality,MeetMap}. Anthology similarly organizes recorded conversations semantically while preserving access to speakers' original voices \cite{Anthology}. T-Cal \cite{TCal} and Tilda \cite{Tilda} focus on asynchronous conversations to help users understand what people discussed in chat-based conversations, while EchoMind supports sensemaking during complex problem-solving discussions \cite{EchoMind}. Systems such as Rationale \cite{RationaleOnline} and ArguMap \cite{ArguMap} structure conversations around argumentation to make supporting and opposing points explicit, while AMQuestioner extends argument maps with question-driven support for critical thinking \cite{AMQuestioner}. Similarly, CueTip \cite{CueTip} generates contextual conversational cues to increase idea diversity with the conversations.

A second group of systems supports sensemaking across multiple conversations or groups. Groupnamics provides awareness of multiple classroom discussions and their speaking dynamics, while VizGroup surfaces AI-generated events and collaboration metrics across programming groups, and ZoomSense provides infrastructure for exposing real-time meeting activity \cite{Groupnamics,VizGroup,ZoomSense}. Needle visualizes online text-based discussion threads and allows moderators to track measures such as positivity and toxicity \cite{needlingViz}, while instructor-facing interfaces for online classrooms similarly support awareness of student engagement \cite{stuEmoOnlineClass}. NAIRA supports instructors in monitoring group discussions through measures such as silence, individual speaking time, and user--user influence relationships that can help reveal dominating speakers \cite{NAIRA}. CPVis supports post-hoc analysis of multimodal collaborative behavior \cite{CPVis}, while GroupTrackVis supports visual analysis of discussions across different groups \cite{GroupTrackVis}. Dynamite aggregates conversational data from multiple groups together as a statistical plot \cite{Dynamite}. MobileGroupVis, which is particularly close to our setting, supports in-situ visual analysis of multiple ongoing group discussions through a mobile interface \cite{MobileGroupVis}.

While prior systems provide substantial support for making sense of conversational content and interaction across groups, voice characteristics can contain additional information that may not be apparent from transcripts alone. Our work investigates conversation sensemaking that incorporates such acoustic information alongside discussion content and other interaction dynamics. We focus particularly on live CR activities, where instructors need to understand not only what students are discussing, but also aspects of how students are collaborating with one another.

\subsection{Evidence Grounding in Human-AI Sensemaking}

Sensemaking involves moving iteratively between available evidence and higher-level interpretations \cite{pirolli2005sensemaking}. Human-AI systems have supported this movement in different ways. Sensecape enables users to move across multiple levels of abstraction while alternating between information foraging and sensemaking \cite{Sensecape}, while Graphologue transforms LLM-generated information into interactive diagrams that users can further explore through contextual queries \cite{Graphologue}. Other systems more explicitly connect AI-generated abstractions back to their supporting sources. Traceable Texts links phrases in generated summaries to corresponding passages in source documents \cite{TraceableTexts}, while PaperTrail decomposes LLM responses and source documents into claims and evidence to support fine-grained provenance inspection \cite{PaperTrail}. ForSense similarly integrates information gathering and sensemaking during online research while providing machine assistance across different stages of the process \cite{ForSense}. Together, these systems illustrate how AI-generated representations can serve as intermediate structures through which users move between summarized information and more detailed supporting evidence.

Evidence grounding becomes more challenging when the underlying information continues to evolve. AIR allows meeting participants to revisit missed portions of an ongoing meeting through audio, video, transcripts, and shared-workspace information \cite{AIR}, while CatchLive provides evolving summaries and highlight moments that help viewers understand earlier portions of a live stream without completely disengaging from the ongoing content \cite{CatchLive}. These systems demonstrate the value of preserving access to earlier evidence during live activities, but largely consider a single evolving information stream. In concurrent live conversations, inspecting one piece of evidence additionally consumes attention while several other conversations continue to change. \sys{} investigates evidence grounding under this temporal and attentional constraint, supporting instructors in moving between compact AI-generated representations and the transcript and audio evidence underlying multiple simultaneously evolving conversations.

\section{Formative Study}

To better understand these challenges, we developed an early node-based prototype and tested it with three TAs in a simulated live multi-group discussion environment. The prototype was informed by observations of CR activities and the types of signals identified as useful through an initial dashboard prototype deployed in Ethics classes taught by two instructors at our institution (E1 and E2), as well as feedback from semi-structured interviews with three additional instructors (E3, E4, and E5). Based on these formative activities, we subsequently derived a set of design goals for \sys{}. This research was approved by our institution's IRB.

\subsection{Early Node-Based Prototype and Formative Evaluation}

Our early prototype was designed around the idea of making conversational evidence accessible through nodes within an AI-generated semantic structure \cite{Graphologue}. This design was motivated by our observations and interviews with instructors E1--E5, as well as the need to support sensemaking at multiple levels, from maintaining a class-level overview to inspecting more detailed group-level information. All five instructors described having some expectations about the topics that students should engage with during CR activities. E2 and E3 were also interested in understanding where each group stood on a given topic. In addition, E2--E5 emphasized the need to prioritize groups whose interaction patterns might warrant attention, such as groups that remained silent or, in some cases, groups with strongly divided opinions on topics related to personal beliefs. E4 further noted that students who were particularly ``passionate'' about a topic and expressed strong opinions could sometimes make other students hesitate to share their thoughts.

\begin{figure}[t]
    \centering
    \includegraphics[width=0.65\linewidth]{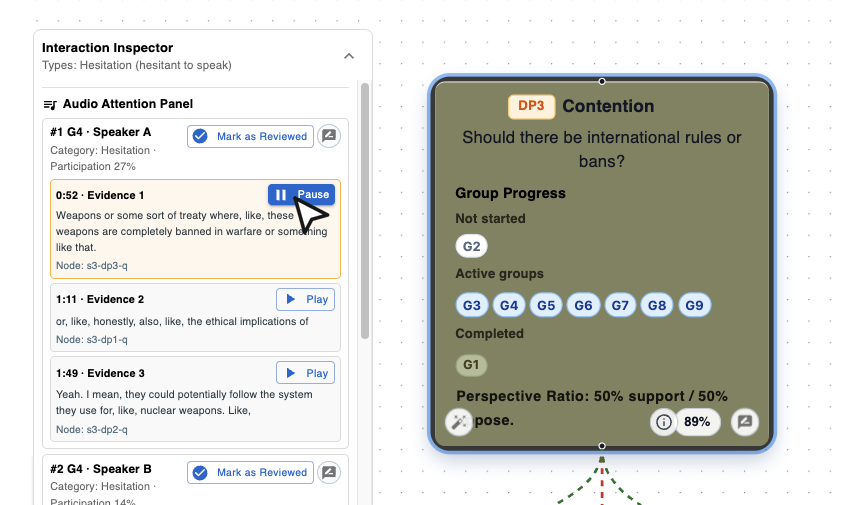}
    \caption{A node-based prototype. This iteration linked node-based representations to audio playback, which informed and motivated \sys{}.}
    \label{fig:nodeVersion}
\end{figure}

The prototype therefore represented expected discussion topics as a semantic node structure that dynamically updated as new information emerged from the groups. Instructors could expand individual nodes to examine which groups had discussed a topic and the specific themes that emerged within each group \cite{Sensecape}. When discussion content deviated from the expected structure, the system generated a new node connected to the most semantically related existing node, allowing unanticipated themes to be incorporated while preserving meaningful relationships among topics \cite{Orality,Graphologue}. As discussions unfolded, the interface represented conversational progress through proportional color filling \cite{conversaionProgressGuide}, with perspective-encoded colors indicating whether groups expressed agreement or disagreement with a topic \cite{RationaleOnline}. Each node also provided access to the evidence underlying the system's interpretation through associated transcript and audio snippets, including audio evidence for interactional signals that may depend on acoustic characteristics of speech \cite{skinner2016another,tan2018analysis} (Fig.~\ref{fig:nodeVersion}).

We conducted a preliminary formative evaluation of this prototype to examine how users foraged for evidence during live multi-group sensemaking. We recruited three TAs (24M, 26F, 28M) with prior experience teaching discussion-based courses and asked them to monitor a simulated live classroom environment constructed from a 10-minute playback of one of the discussion datasets collected from the classes of E1 and E2. Across the three sessions, the classes involved 85, 35, and 34 students, forming 23, 9, and 9 groups, respectively (Appendix~\ref{appendix:dataCollection}). Before the task, participants completed a system tutorial covering all available features using one of the other two datasets as a warm-up dataset. After the main task, we conducted semi-structured interviews to better understand participants' decision-making processes and gather feedback on the system.

Although participants frequently interacted with textual evidence, such as transcripts, none of the three participants used the audio playback features during the monitoring task. In follow-up interviews, all three explained that short audio snippets lacked sufficient surrounding context to make the evidence easy to interpret, particularly for cues such as hesitation. Under the limited attention available during live monitoring, participants instead prioritized information that could be verified more quickly, such as groups that were silent or groups expressing distinctive perspectives on a topic. These observations suggested that simply providing access to short snippets of audio evidence within a typical AI-assisted sensemaking interface was insufficient. Users also needed stronger contextual and temporal support to decide when deeper evidence inspection was worth the associated attentional cost. In addition, all three participants wanted to understand whether an interactional signal was an isolated occurrence or part of a pattern that continued or recurred over time.

\subsection{Design Goals}

Sensemaking involves comparing and contrasting new information with existing knowledge \cite{Sensecape, pirolli2005sensemaking}. Thus, based on our formative study findings, we derived four design goals:

\begin{itemize}
    \item \textbf{DG1: Facilitate interpretation of conversational evidence over time.} The interface should preserve how interactional signals unfold over time, helping instructors distinguish isolated moments from recurring or sustained patterns and interpret individual events within their surrounding conversational context.

    \item \textbf{DG2: Enable quick and low-effort access to conversational evidence.} Because multiple discussions continue to evolve in parallel, the interface should minimize the interaction cost required to access underlying evidence.

    \item \textbf{DG3: Support cross-group sensemaking through comparison.} The interface should present group-level information in a visually comparable form so instructors can quickly identify similarities and differences across conversations.

    \item \textbf{DG4: Guide attention at scale.} The system should help instructors allocate their limited attention across many concurrent discussions by surfacing relevant moments that may require additional inspection.
\end{itemize}

\section{System Overview}

We present \sys{}. Based on Pirolli and Card's sensemaking model \cite{pirolli2005sensemaking}, \sys{} focuses on the information-foraging loop, in which users search through available information, inspect potentially relevant items, and gather evidence for subsequent sensemaking.

While the main focus of this paper is the instructor interface, the \sys{} workflow begins with processing student discussion data in real time (Fig.~\ref{fig:teaser}). The teaching team sets up and shares a link to the student interface, where students can join anonymously and enter their Zoom breakout room number. The system then groups students who enter the same room number. Students can subsequently begin their discussion, and their audio is processed through the real-time collaborative reasoning analysis pipeline described in Section~\ref{sec:cranalysis}.

The processed information from each student group is then populated on the \sys{} instructor dashboard. Guided by DG1--DG4, the interface is designed to support evidence foraging across concurrent group discussions. We describe the instructor interface in Section~\ref{sec:sysinterface}.

\section{Real-Time Collaborative Reasoning Analysis}
\label{sec:cranalysis}

\sys{} analyzes ongoing group discussions using six collaboration-relevant cues: \textit{Disagreement}, \textit{Opinion}, \textit{Coordination}, \textit{Hesitation}, \textit{Uncertainty}, and \textit{Interruption}, informed by prior literature on collaborative discourse and speech (Table~\ref{tab:crQuality}). To support real-time analysis, \sys{} transcribes incoming speech using Deepgram \cite{Deepgram} and tracks semantic topics using vector embeddings and cosine similarity \cite{Dynamite, MobileGroupVis}. Audio is captured through the browser using the HTML5 \texttt{MediaRecorder} API and anonymized while preserving voice-quality characteristics relevant to conversational analysis \cite{anonVoiceDialogue, audioManiEffect}.

Because these cues arise from different aspects of conversation, we organize them into three categories according to their primary source of evidence. \textit{Lexical cues} are derived from the semantic content of what participants say, \textit{acoustic cues} from how speech is delivered, and \textit{speaker-timing cues} from the timing and relationships between speaker turns.

\begin{table*}[t]
\centering
\caption{Literature-informed characterizations of the six collaboration-relevant cues represented in \sys{}, broadly grouped into collaboration-related and reasoning-related characteristics.}
\label{tab:crQuality}
\small
\setlength{\tabcolsep}{6pt}
\renewcommand{\arraystretch}{1.15}
\begin{tabular}{
    >{\raggedright\arraybackslash}p{0.18\textwidth}
    >{\raggedright\arraybackslash}p{0.40\textwidth}
    >{\raggedright\arraybackslash}p{0.32\textwidth}}
\toprule
\textbf{Cue} &
\textbf{Literature-Informed Characterization} &
\textbf{Pattern That May Warrant Attention} \\
\midrule

\multicolumn{3}{l}{\textit{\textbf{Collaboration-Related Characteristics}}} \\
\midrule

\textbf{Coordination}
&
Collaborative interaction involves managing participation and joint progress, including inviting contributions, coordinating activity, and transitioning between discussion points \cite{kumpulainen2003interpersonal, groupCommAnalysis, LLMClassDebateCHI}.
&
Coordination becomes limited or unclear, with students taking little initiative to manage the group's discussion.
\\

\textbf{Hesitation}
&
Hesitation in spontaneous speech can be expressed through silent pauses and filled pauses, including ``um'' and ``uh'' \cite{maclay1959hesitation, clark2002using}.
&
Students may exhibit disfluencies when sharing their thoughts on a case study with peers.
\\

\textbf{Uncertainty}
&
Prosody and voice quality can convey expressive and metacognitive states, with rising or high utterance intonation associated with perceived uncertainty or feeling of knowing \cite{prosodyStorytelling, voiceQualityInPerception, brennan1995feeling, swerts2005audiovisual}.
&
Uncertainty may indicate a need for clarification or response to help move the conversation forward.
\\

\textbf{Interruption}
&
Interruptions can emerge through overlapping speaker turns and changes in how participants take the conversational floor \cite{tan2018analysis}.
&
Interruptions may suppress the exchange of ideas, allowing some students to dominate the group and limiting exposure to diverse perspectives \cite{anEthicsPrimer}.
\\

\midrule
\multicolumn{3}{l}{\textit{\textbf{Reasoning-Related Characteristics}}} \\
\midrule

\textbf{Opinion}
&
Collaborative reasoning involves introducing substantive claims, information, and new or diverse perspectives that can be further developed through evidence and explanation \cite{DiscussionTracker, MobileGroupVis, groupCommAnalysis, minorityVoicesDevilsAdvocate}.
&
Group discussions that do not introduce new ideas may become stagnant and explore less content \cite{embededEthics}.
\\

\textbf{Disagreement}
&
Questioning or challenging prior ideas can surface alternative perspectives and prompt further justification or critical examination \cite{DiscussionTracker, anEthicsPrimer, LLMDevilsAdvocate}.
&
Group discussions without disagreement may not sufficiently surface contrasting perspectives, which can support the practice of reasoning skills \cite{DiscussionTracker,LLMClassDebateCHI}.
\\

\bottomrule
\end{tabular}
\end{table*}

\subsection{Lexical Cues}

Based on prior literature (Table~\ref{tab:crQuality}), \sys{} operationalizes three lexical cues.

\subsubsection{Opinion}

We operationalize an \textit{Opinion Cue} as a speaker introducing a claim, factual or reference information, or another substantive idea that is new relative to ideas expressed before. For example, an \textit{Opinion Cue} may be identified when a student says, ``I think if you put your artwork on the public Internet, well, definitely it's something that you have to expect robots will crawl data on them.''

\subsubsection{Disagreement}

We operationalize a \textit{Disagreement Cue} as a speaker questioning, challenging, or expressing disagreement with an idea introduced in a prior speaker turn. Here, a \textit{turn} refers to a continuous contribution from one speaker before another speaker takes the floor. For example, the following response would be treated as a \textit{Disagreement Cue}: ``That's a fair point. However, what about students who were trying to do a school project? Should the same law be applied in that case?'' Within a single turn, a speaker may disagree with multiple previous turns, but the same prior turn is not counted as a disagreement more than once.

\subsubsection{Coordination}

We operationalize a \textit{Coordination Cue} as meta-discourse through which students manage or direct the group discussion, such as introducing a discussion question, inviting others to contribute, or proposing a transition between discussion points. Examples include ``Let's move to the next question'' and ``What do you guys think about this topic?''

\subsection{Acoustic Cues}

Compared to lexical cues, acoustic cues are more tacit and require human interpretation, primarily of acoustic characteristics and, in some cases, in combination with lexical information, to understand the underlying group dynamics. To operationalize these concepts, we refined the literature-informed characterizations through an iterative annotation process using our collected datasets. In this process, we examined two datasets comprising 18 groups and 69 students. With approximately 10 minutes of discussion per group, these datasets contained roughly 180 minutes of conversational audio.

Two researchers first developed annotation rubrics for \textit{Hesitation} and \textit{Uncertainty} based on the literature-informed characterizations in Table~\ref{tab:crQuality}. We independently annotated the first group from one of the datasets, discussed and resolved differences, and updated the rubrics based on our consensus. We then repeated this process with a second group from the datasets, again resolving differences and refining the rubrics. We supported this process with an annotation interface visualizing the waveform and pitch contour of each utterance alongside its audio. For each group, the researchers annotated the conversation chronologically, starting from the beginning of the session and proceeding to the end of the 10-minute discussion. After finalizing the rubrics, the researchers independently annotated the remaining groups, using the same methodology, and computed Cohen's $\kappa$ for each cue.

\subsubsection{Hesitation}

We operationalize a \textit{Hesitation Cue} as an utterance containing multiple disfluencies. Disfluencies include verbal fillers such as ``um'' or ``uh'', pauses within or between sentences, and prolonged speech segments. Pauses within a sentence had to last at least one second, while pauses between sentences had to last at least 1.5 seconds. Prolonged speech segments also had to last at least 1.5 seconds. A single disfluency alone was not considered sufficient, as such disfluencies commonly occur while speakers are finding the right words and do not necessarily indicate hesitation about contributing an idea. Therefore, the first detected disfluency was not labeled. When a second disfluency was encountered within the same turn, a \textit{Hesitation Cue} was labeled and the count would be reset to zero. As a result, a single speaker turn could contain multiple \textit{Hesitation Cues}. The annotation therefore focused on more sustained or repeated disfluency patterns that provide stronger evidence of hesitant delivery. The resulting Cohen's $\kappa$ was 0.623.

\subsubsection{Uncertainty}

We operationalize an \textit{Uncertainty Cue} as an utterance exhibiting a salient upward excursion in fundamental frequency ($F_0$). To account for differences in speakers' pitch without requiring a prior speaker profile, we measure pitch relative to the utterance itself. Specifically, we compute the median $F_0$ across voiced frames and compare it with the 95th-percentile $F_0$, expressing this difference in semitones. Based on refinement using our annotated development data, utterances exhibiting an upward pitch excursion of at least 4 semitones are considered candidate \textit{Uncertainty Cues}. The resulting Cohen's $\kappa$ was 0.634.

\subsection{Speaker-Timing Cues}

Speaker-timing cues characterize interactional events derived directly from the temporal relationship between speaker turns. \sys{} currently represents \textit{Interruption}.

\subsubsection{Interruption}

We operationalize an \textit{Interruption Cue} as an overlap between two speaker turns, identified when a new speaker begins speaking before the preceding speaker's turn has ended.

\subsection{Real-Time Computation Pipeline}

We implemented a real-time processing pipeline that detects signals from incoming transcript and audio streams. Because the signals rely on different forms of conversational evidence, \sys{} uses separate computation pipelines for lexical, acoustic, and speaker-timing signals. 

\begin{figure}
    \centering
    \includegraphics[width=1\linewidth]{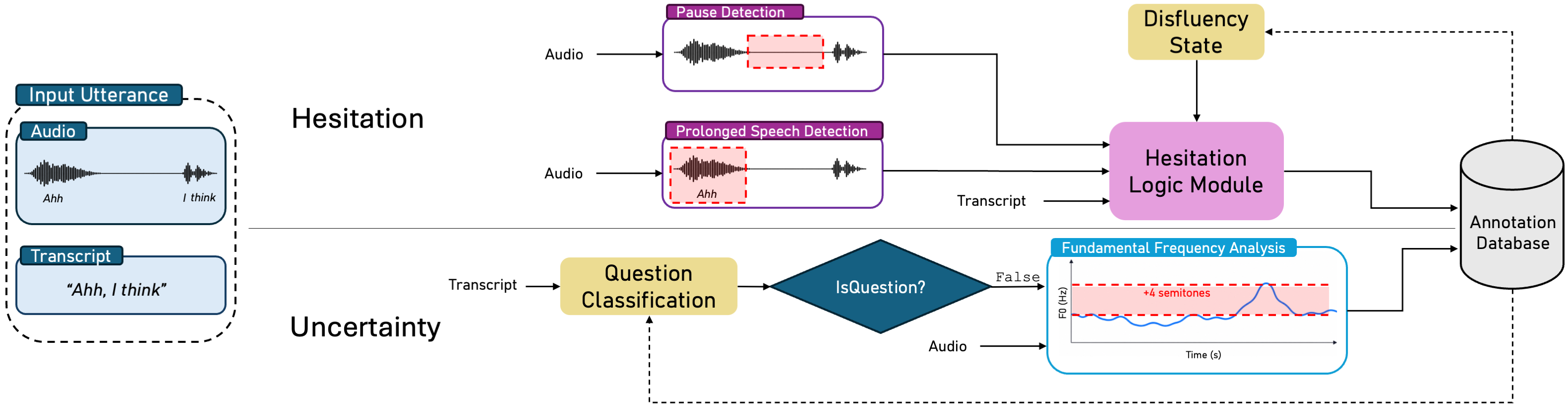}
    \caption{Overview of the acoustic cue computation pipeline in \sys{}, which processes each input utterance using its audio and transcript components. The top section detects \textit{Hesitation Cues} by combining filler-word, pause, and prolonged-speech detection with a disfluency state accumulated across utterances within the same speaker turn, while the bottom section detects candidate \textit{Uncertainty Cues} by filtering questions and analyzing fundamental-frequency differences in the incoming audio. Detected cues are timestamped and stored in the annotation database before being presented in the instructor interface.}
    \label{fig:acousticPipeline}
\end{figure}

\subsubsection{Lexical Signals}

To detect lexical signals, we developed a few-shot in-context classification pipeline using GPT-4o-mini through the OpenAI API. For \textit{Opinion} and \textit{Disagreement Cues}, we constructed the classification prompts using labeled examples from the Discussion Tracker corpus, which includes collaborative discourse behaviors such as introducing new ideas and challenging or disagreeing with prior contributions \cite{DiscussionTracker}. For \textit{Coordination Cues}, we provided researcher-labeled examples from our own classroom corpus. The lexical cues are computed at different units of conversation. \textit{Opinion} and \textit{Coordination Cues} are classified at the utterance level, allowing each incoming utterance to be evaluated as it arrives. In contrast, \textit{Disagreement Cues} are classified at the speaker-turn level. Because a speaker turn may consist of multiple incoming utterances, \sys{} accumulates consecutive utterances from the same speaker to construct the turn before classification. The current contribution and preceding conversational context are then provided to GPT-4o-mini together with the corresponding few-shot examples and cue definition. The model returns the predicted cue type together with its corresponding transcript evidence and timestamp.

\subsubsection{Acoustic Signals}

As shown in Fig.~\ref{fig:acousticPipeline}, the acoustic pipeline processes each incoming utterance using its audio and transcript components. We implemented the pipeline in Python using WebRTC Voice Activity Detection (VAD) for identifying pauses and \texttt{librosa} for fundamental-frequency ($F_0$) analysis. For \textit{Hesitation Cues}, the pipeline detects qualifying pauses and prolonged speech from the audio and filler words from the transcript. Because a speaker turn may contain multiple utterances, the system maintains a disfluency state across consecutive utterances within the same turn. The Hesitation Logic Module applies the thresholds defined above to the current and accumulated disfluencies, labels a \textit{Hesitation Cue} when the criteria are met, resets the disfluency state, and stores the resulting cue in the annotation database. For \textit{Uncertainty Cues}, the pipeline estimates the $F_0$ contour of each utterance using \texttt{librosa}'s \texttt{pyin} algorithm. It computes the median and 95th-percentile $F_0$ across voiced frames and converts their difference to semitones. An utterance with a difference of at least four semitones is marked as a candidate \textit{Uncertainty Cue} if the corresponding transcript is classified as a non-question.

\subsubsection{Speaker-Timing Signals}

To detect \textit{Interruption Cues}, \sys{} compares the start and end timestamps of consecutive speaker turns. An interruption is marked whenever a speaker begins a turn before the preceding speaker's turn has ended. The resulting event is timestamped at the beginning of the overlapping turn.

\subsection{Technical Evaluation}
\label{sec:technicalEvaluation}

We conducted a technical evaluation focused on \textit{Hesitation} and \textit{Uncertainty}, whose detection depends on the acoustic operationalizations described above. We did not separately evaluate the lexical cues, whose underlying discourse constructs and automated classification have been extensively studied in prior work. For instance, MobileGroupVis~\cite{MobileGroupVis} evaluated its analysis pipeline through case-study deployments, demonstrating its feasibility for real-time classroom use. We also excluded \textit{Interruption} because it is computed directly from overlapping speaker-turn timestamps.

\subsubsection{Acoustic Cue Detection Accuracy}

Building on the dataset described above, we used the first two of the 18 groups to develop and refine the annotation rubric and computational thresholds. From the remaining groups, we sampled ten groups across the two 10-minute discussion sessions, yielding 1,264 utterances for evaluation. Three researchers independently annotated each utterance for \textit{Hesitation} and \textit{Uncertainty}, and the final label was determined through majority voting.

We compared \sys{}'s predictions against these gold-standard annotations. For \textit{Hesitation}, the pipeline achieved a precision of 0.801, recall of 0.979, and F1 score of 0.881. For \textit{Uncertainty}, it achieved a precision of 0.773, recall of 1.000, and F1 score of 0.872. Errors were concentrated primarily in false positives. For hesitation, these sometimes occurred when speakers with relatively lower English fluency frequently used filler words and pauses throughout their speech. For uncertainty, false positives included speakers who habitually raised their pitch near the end of utterances as part of their speaking style.

\subsubsection{Real-Time Processing Feasibility}

Because \sys{} is designed for live classroom monitoring, we also measured the processing time required for acoustic cue detection on a server with 1~vCPU and 2~GB of memory. \textit{Hesitation} detection required a median of 132~ms (IQR~=~114--152~ms) across 1,264 utterances, while \textit{Uncertainty} detection required a median of 250~ms (IQR~=~201--298~ms).

\subsubsection{In-Class Deployment}

We further deployed \sys{} in two online class sessions taught by E1 with IRB approval. In each session, 12 students were divided into four groups and participated in a 10-minute CR activity, with the system hosted on a server with 1~vCPU and 2~GB of memory. Across the two deployments, \sys{} generated 283 cue annotations, including 21 \textit{Disagreement}, 104 \textit{Opinion}, 16 \textit{Coordination}, 85 \textit{Hesitation}, 43 \textit{Uncertainty}, and 14 \textit{Interruption} cues. E1 used the instructor dashboard to navigate these cues while monitoring the ongoing group discussions in real time. Together, these deployments demonstrate that \sys{} can process concurrent classroom conversations end to end and make the resulting conversational evidence available through the instructor-facing interface in real time.

\section{The \sys{} System}
\label{sec:sysinterface}




The main instructor interface of \sys{} is a web-based dashboard designed to support evidence-grounded sensemaking across multiple concurrent CR discussions. The interface provides temporally aligned representations that help instructors maintain awareness across groups, identify conversational moments that warrant closer inspection, and access the transcript and audio evidence underlying those representations.

\subsection{Temporal Signal Density Visualization}

The primary visualization in \sys{} is a temporal signal density plot that summarizes the distribution of selected collaboration-relevant cues along each group's conversational timeline. \sys{} aggregates nearby signals into a continuous density representation. Regions containing more closely occurring signals form higher peaks, while isolated signals appear as localized changes. 

This representation provides temporal context for interpreting cues (DG1), allowing instructors to distinguish sustained or recurring patterns from one-off events. By making these differences visually apparent at a glance, the density plot also supports rapid evidence inspection under live time constraints (DG3). The density plot can also be expanded into multiple cue-specific tracks, allowing instructors to view the individual collaboration-relevant cues underlying the aggregated timeline. 

\begin{figure}
    \centering
    \includegraphics[width=0.8\linewidth]{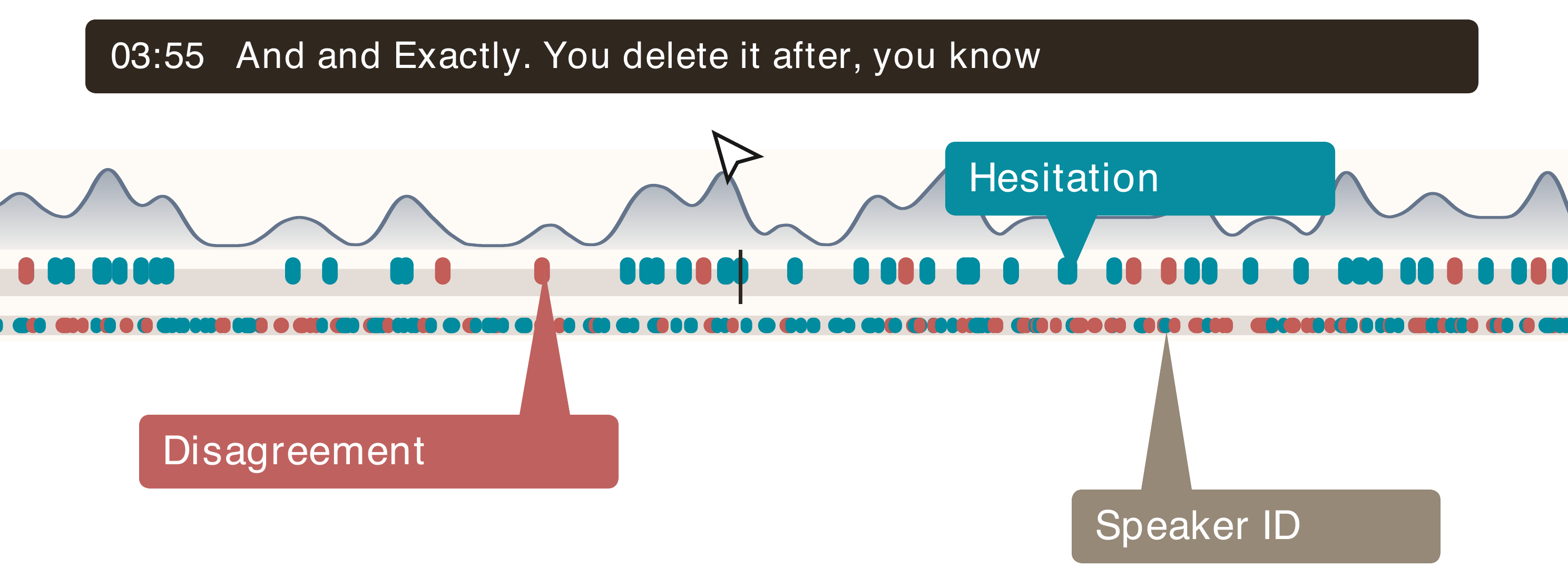}
    \caption{The temporal signal density visualization summarizes how selected collaboration-relevant cues are distributed over time. Individual signal annotations and their corresponding evidence are accessible for further inspection.}
    \label{fig:signalDensity}
\end{figure}

\subsubsection{Representing Temporal Patterns}

The density plot is computed from the collaboration-relevant cues currently selected by the instructor. Signals occurring within the same window are aggregated, and the resulting counts are normalized relative to the highest signal concentration observed for that group. The vertical amplitude therefore represents relative cue concentration within each 7.5-second interval. We selected this short temporal window to preserve local conversational variation while still aggregating nearby events into interpretable patterns, consistent with prior conversational analyses that use similarly sized windows to characterize local interaction dynamics \cite{duran2017conversing}.

As signals accumulate over time, higher-density regions indicate intervals containing greater concentrations of detected signals, while isolated events produce more localized changes. This allows instructors to distinguish one-off occurrences from recurring, clustered, or sustained patterns (DG1). Peaks therefore serve as cues to conversational regions that instructors may choose to inspect in greater detail.

\subsubsection{Supporting Cross-Group Comparison}

The same density encoding is applied consistently across all group tiles. Instructors can therefore compare the temporal distribution of collaboration-relevant cues across groups (DG3). 

\subsubsection{Accessing Underlying Evidence}

The density plot is directly linked to the conversational evidence underlying each region. \sys{} supports evidence navigation across multiple levels of detail, from the aggregated density representation to individual collaboration-relevant cues, transcript segments, and source audio. This intermediate representation layer helps instructors move between high-level patterns and lower-level conversational evidence without immediately committing attention to raw audio, which is designed for interactional cues that unfold continuously over time and may be difficult to capture through textual abstraction alone.

Hovering over the visualization reveals the corresponding transcript. Clicking a point along the timeline moves the audio playhead to the corresponding moment, allowing instructors to listen to the raw audio when additional acoustic context is needed (DG2). Because earlier portions of the timeline remain available as new discussion data arrives, instructors can also revisit previously detected moments after shifting their attention to other groups.

\subsection{Monitoring Dashboard Overview}

To support monitoring of multiple concurrent group discussions, the interface is organized into three primary areas: a Group Monitoring Gallery, an Signal Control Panel, and a Focused Group Panel.

\begin{figure*}[t]
    \centering
    \includegraphics[width=0.75\linewidth]{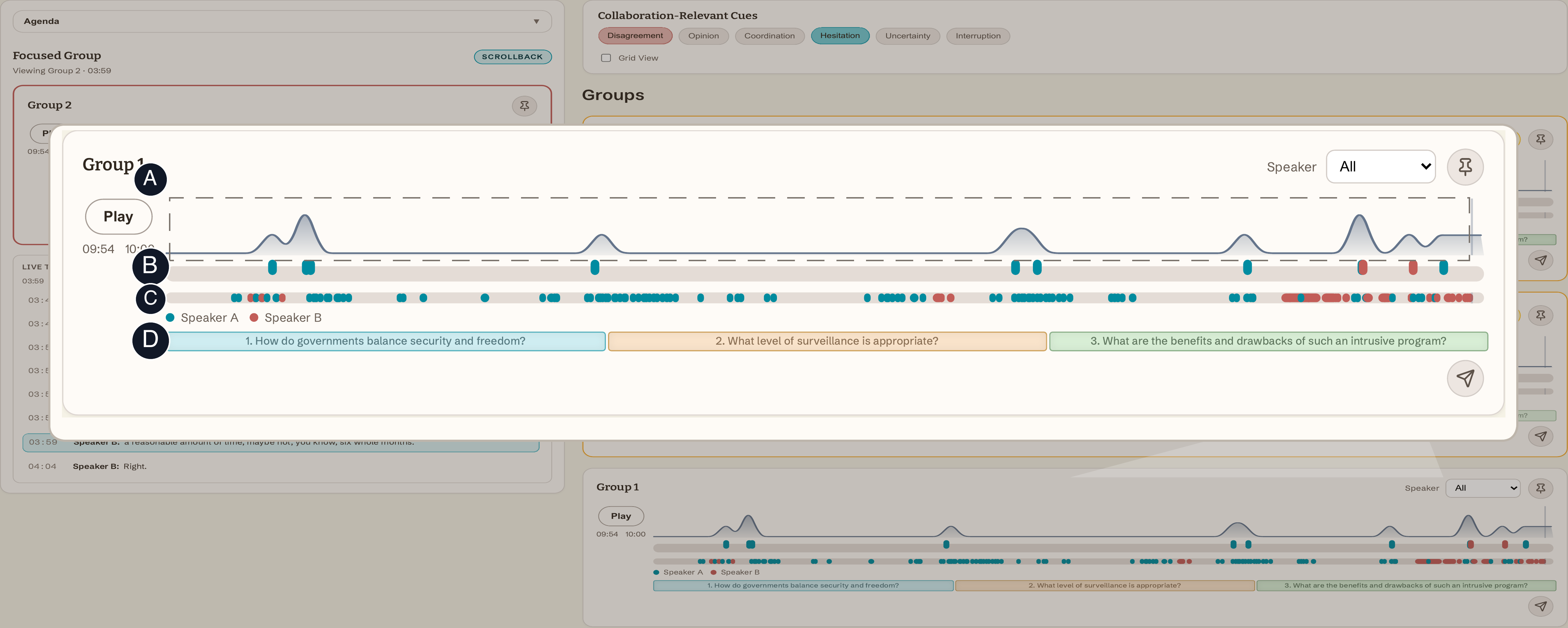}
    \caption{List View of the group monitoring dashboard showing (A) the temporal signal density visualization, (B) individual interaction-signal annotations, (C) speaker distribution, and (D) content coverage.}
    \label{fig:listviewfig}
\end{figure*}

\subsubsection{Group Monitoring Gallery}

The Gallery (Fig.~\ref{fig:teaser}E) provides an overview of all parallel discussions. To support varying class sizes and screen ratios, \sys{} provides two layout modes: a Grid View for monitoring multiple groups simultaneously (e.g., a 3$\times$3 grid for 9 groups), and a List View that displays groups as full-width rows to provide greater temporal resolution for each timeline.

\subsubsection{Cue Control Panel}

The Cue Control Panel allows instructors to select which collaboration-relevant cues are displayed across the Gallery. Enabling or disabling a cue type updates both its individual annotations and its inclusion to the temporal density visualization. This allows instructors to focus the interface on specific interactional behaviors while retaining the same underlying conversational timeline.

\subsubsection{Focused Group Panel}

Selecting a group populates the Focused Group Panel with a synchronized audio player and time-stamped transcript (Fig.~\ref{fig:teaser}F). Instructors can navigate to any point in the conversation through the timeline or transcript and inspect the corresponding audio. The panel remains fixed on the left side of the interface, allowing instructors to inspect a selected group while maintaining visibility of the other ongoing group discussions on the right. During an active session, instructors can listen to live audio streamed directly from the groups to support real-time discussion management. For historical sessions, however, audio is retrieved from the database and therefore presented in anonymized form.

\subsection{Discussion Content Visualization and Tracking}

\begin{figure}
    \centering
    \includegraphics[width=0.8\linewidth]{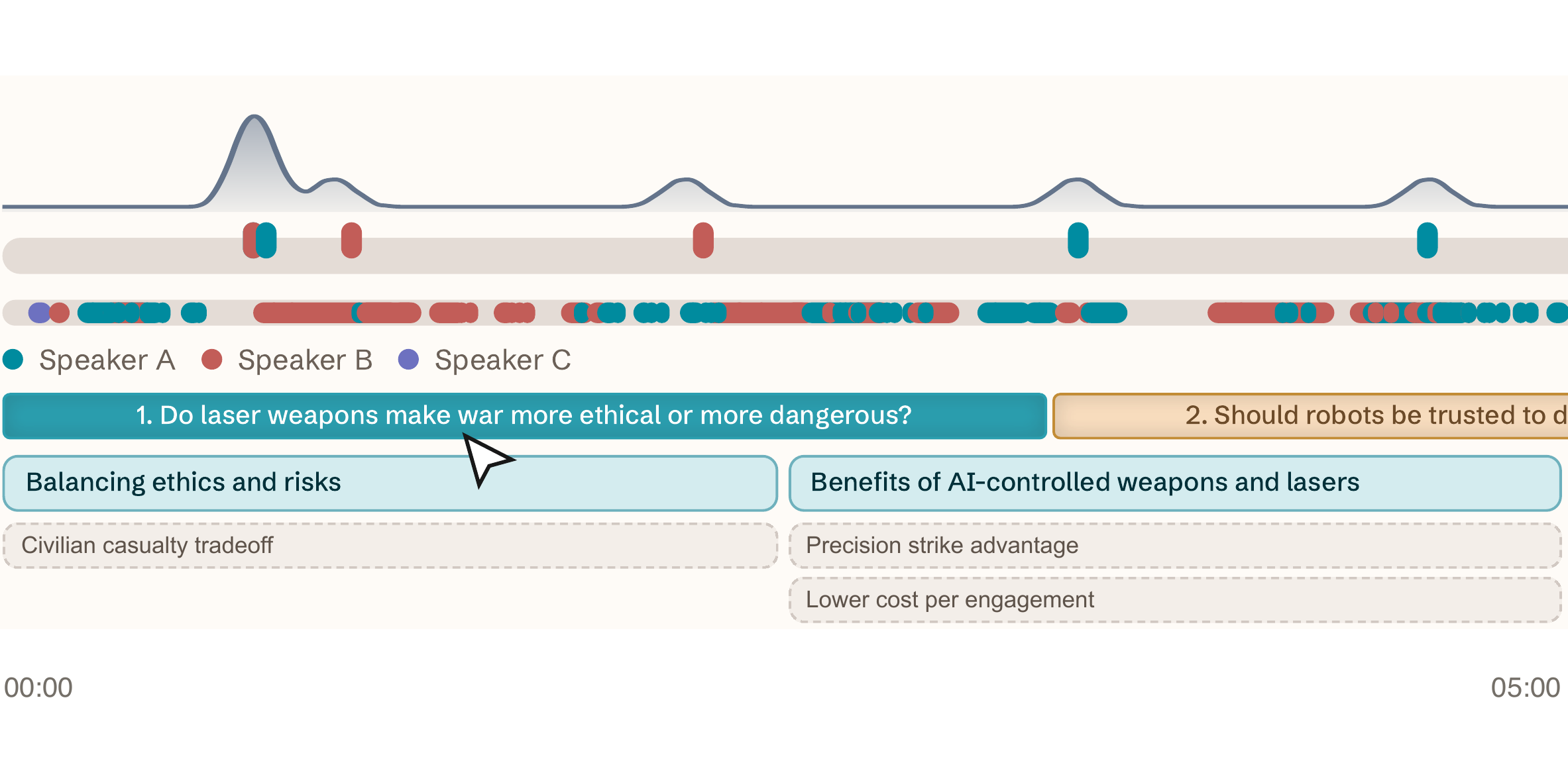}
    \caption{Discussion topics are represented along the group timeline to show content coverage over time. Hovering reveals the corresponding topic and discussion content.}
    \label{fig:expectedTopics}
\end{figure}

\sys{} visualizes the topics discussed by each group using a Content Coverage Bar aligned with the same conversational timeline. Topic assignments produced by the semantic topic-tracking pipeline described in the previous section are mapped to temporal regions showing when each topic was discussed.

Instructors can hover over the bar to inspect the corresponding topic and conversational content. Because all groups follow the same topic structure and each discussion point is allocated the same duration, the visualization supports direct comparison of content progression across groups and helps identify groups that have not yet addressed expected discussion points. Aligning topic information with collaboration-relevant cues and speaker activity also allows instructors to examine how discussion content and interactional patterns evolve together over time.

\subsection{Attention Marking}

To support monitoring at classroom scale, instructors can define interactional patterns or discussion states of interest, and \sys{} marks groups that match these conditions for closer attention (DG4). Instructors configure these conditions through a setup interface, with default criteria including prolonged silence and lagging content coverage. They can also define patterns involving collaboration-relevant cues either by specifying parameters manually or by brushing over an example segment of a group timeline. In the latter case, the system extracts the characteristics of the selected pattern and identifies similar occurrences across other groups (DG4).

\subsection{Dynamic Tile Swapping}

\begin{figure}
    \centering
    \includegraphics[width=0.75\linewidth]{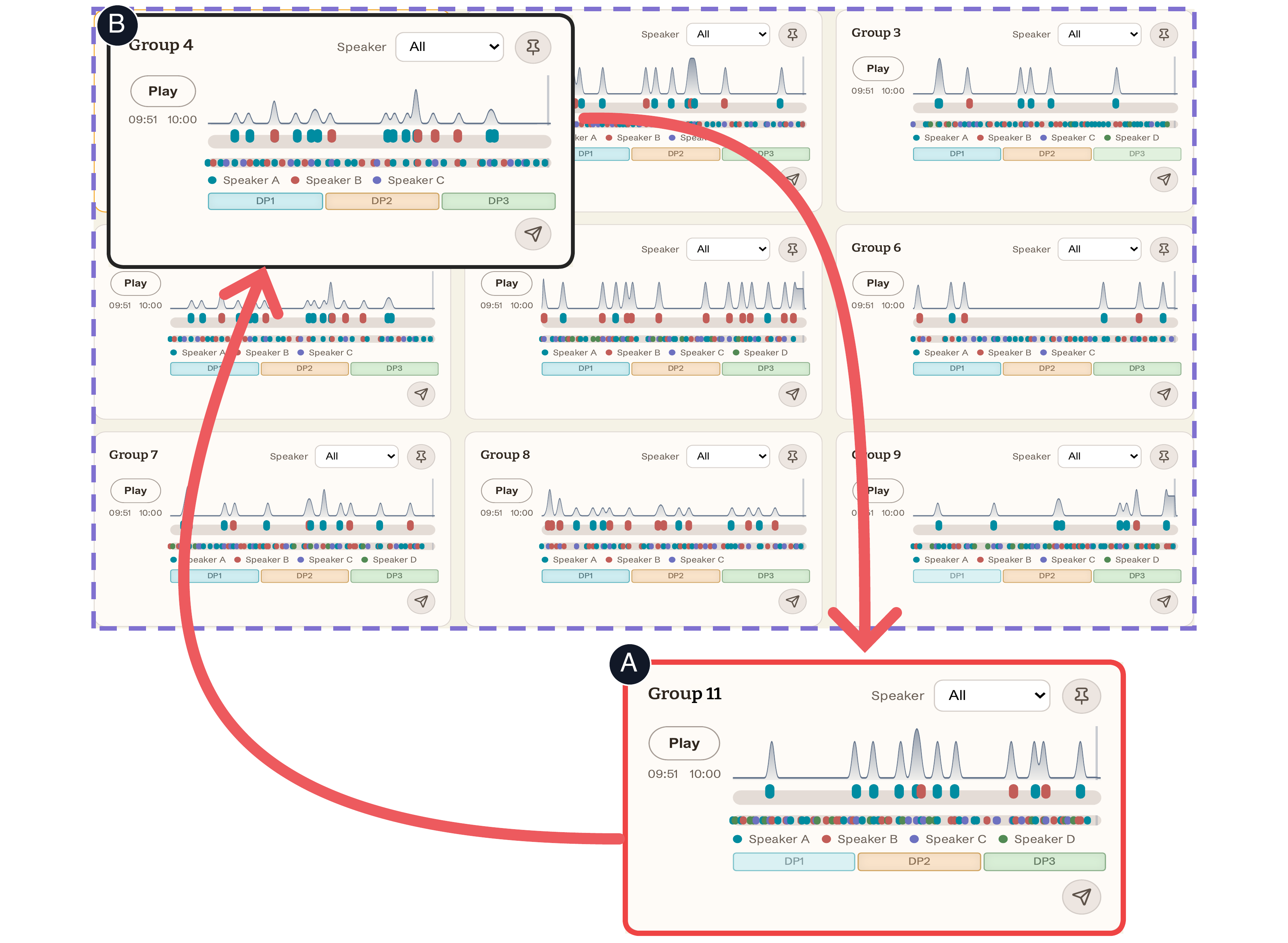}
    \caption{Dynamic Tile Swapping brings higher-priority groups into the visible viewport. When an off-screen group (A, Group 11) is prioritized, the system swaps it with a lower-priority group currently in view (B, Group 4).}
    \label{fig:dynamicTileSwapping}
\end{figure}

When the number of concurrent groups exceeds the available viewport, \sys{} uses dynamic tile swapping to bring groups that need attention (marked by attention marking) into view (DG4; Fig.~\ref{fig:dynamicTileSwapping}). Tile positions are updated every 10 seconds~\cite{Codeopticon}. To avoid disrupting ongoing inspection, tiles under active cursor interaction are excluded from swapping.

\subsection{Pin Groups}

Instructors can pin groups to keep them visible in the Gallery and prevent them from being replaced through dynamic tile swapping. This allows instructors to preserve groups they want to continue monitoring or revisit while the rest of the Gallery adapts to changing priorities across conversations.

\subsection{Other Features}

\sys{} provides additional features for acting on information identified during monitoring.

\subsubsection{Message to Group}

\begin{figure}
    \centering
    \includegraphics[width=0.65\linewidth]{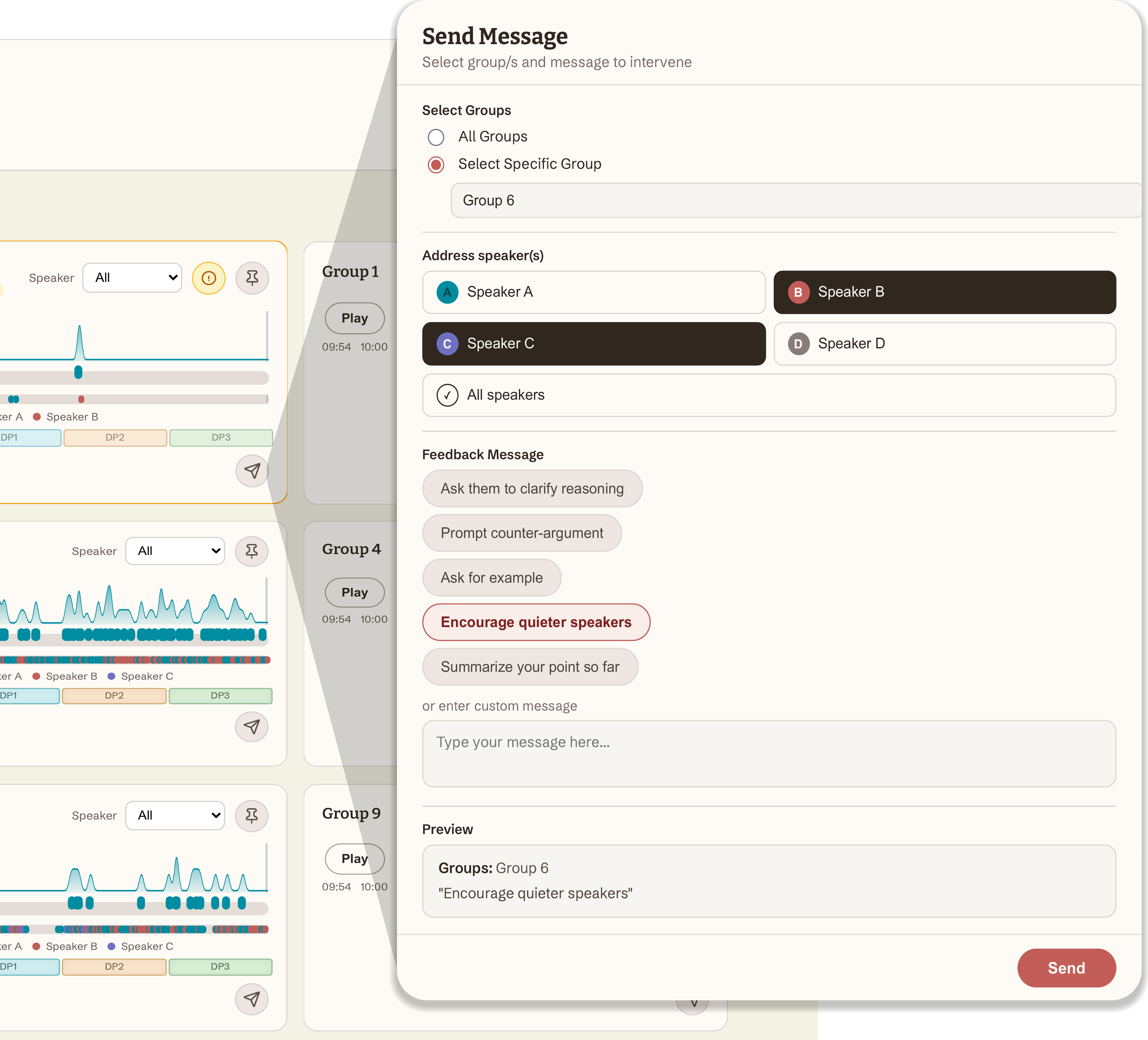}
    \caption{The Message to Group feature allows instructors to send feedback directly to a group. The interface also provides suggested messages that instructors can review before sending.}
    \label{fig:msgToGroup}
\end{figure}

The \textit{Message to Group} feature allows instructors to send text feedback directly to a selected group without leaving the dashboard. The interface can also provide suggested messages that instructors may review and modify before sending. After an instructor addresses an attention marking, the corresponding marking can be cleared from the Gallery.

\subsubsection{Automatic Messaging for Group Progress}

For predefined content-progress conditions, instructors can optionally enable automatic messaging. This is intended for cases where the same easily identifiable condition occurs across many groups, such as several groups falling behind the expected topic progression. When automatic messaging is enabled for a condition, the corresponding attention markings can be suppressed so that the Gallery continues to prioritize situations requiring instructor inspection.



\section{Evaluation}

We conducted a mixed-design in-lab user study to investigate how \sys{} supports evidence-grounded sensemaking across concurrent classroom CR discussions. Our evaluation was guided by two research questions: \textbf{RQ1:} \textit{How does \sys{} affect users' ability to locate and inspect relevant conversational evidence during live sensemaking across concurrent discussions?} and \textbf{RQ2:} \textit{What evidence-foraging and grounding strategies do users adopt while making sense of multiple concurrently evolving conversations with \sys{}?}

We designed the study in two complementary parts. Part 1 focused on objective monitoring performance, where we measured the accuracy and time required to locate conversational evidence. Part 2 involved a continuous simulated sensemaking task based on concurrent 10-minute CR discussions, where we analyzed participants' evidence-inspection behaviors and post-task survey responses.

\subsection{Methodology}

We recruited 20 participants (10 male, 8 female, and 2 of other genders), including 3 instructors and 17 TAs, all of whom had prior teaching experience and experience facilitating group discussions in the course context. Participants (P1--P20) were recruited through internal mailing lists and personal contacts. We compared three interface conditions, all of which retained access to the underlying conversational audio:

\begin{itemize}
    \item \textbf{\sys{} (A).} The full system, including representations of discussion content and collaboration-relevant cues.
    \item \textbf{\sys{} Discussion Content (B).} An ablated version of \sys{} that retained representations of what groups discussed and how their discussion progressed, but removed the collaboration-relevant cue visualizations.
    \item \textbf{Transcript Baseline (C).} A transcript interface that supported scrolling, keyword search, and access to corresponding audio.
\end{itemize}

We selected this study setting to isolate the contribution of collaboration-relevant cue features, which are central to the contribution of this paper, while retaining a discussion-content component that reflects the semantic monitoring support provided by prior systems~\cite{MobileGroupVis, GroupTrackVis, Dynamite}.

Participants were assigned to one of two condition pairings. One group compared the Transcript Baseline (C) with the full \sys{} condition (A), while the other compared the Transcript Baseline (C) with the \sys{} Discussion Content condition (B). In Part 2, participants used either the full \sys{} condition or the Discussion Content condition according to their assigned study group. Participants interacted with recorded classroom CR discussion sessions, each containing multiple parallel group discussions lasting approximately ten minutes.

Each study session was audio- and screen-recorded, lasted approximately 1.5 hours, and participants received \$25 USD in compensation. The research protocol was approved by our institution's IRB.

\subsection{Part 1: Monitoring Performance}

Part 1 was designed to provide a controlled measure of evidence-location performance for RQ1. Participants completed a series of quiz questions using two prototype conditions. The quizzes included questions about discussion content and collaboration-relevant behaviors. Before participants used each condition, we provided a walkthrough of the relevant features using a warm-up dataset to ensure familiarity. We counterbalanced both the order of the prototype conditions and the datasets used across participants.

We measured participants' performance using completion time, precision, and recall. Together, these measures captured how efficiently and accurately participants could forage for relevant conversational evidence for live sensemaking.

\subsection{Part 2: Simulated Real-time Monitoring and User Feedback}

\begin{figure}
    \centering
    \includegraphics[width=0.8\linewidth]{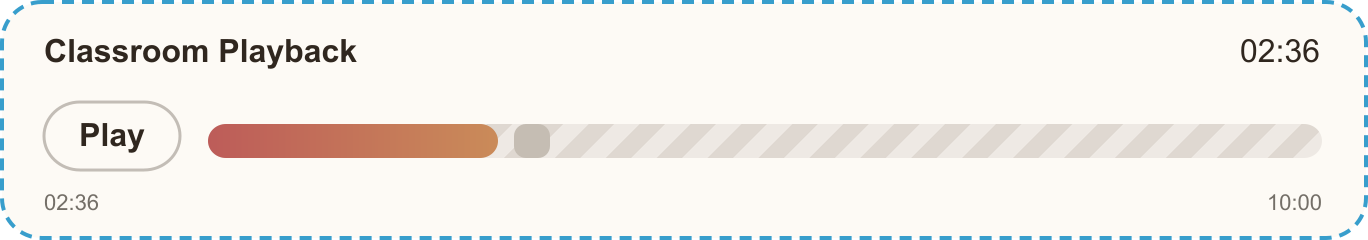}
    \caption{Classroom Playback feature used in the user study, allowing participants to navigate to specific timestamps for answering questions and simulating real-time classroom monitoring.}
    \label{fig:playback}
\end{figure}

Part 2 examined evidence grounding under the continuous attention demands of live classroom monitoring. Unlike the targeted tasks in Part 1, participants monitored a complete 10-minute classroom discussion in real time while multiple group conversations evolved in parallel. Participants used either the full \sys{} condition or the ablated version according to their study assignment. They were also introduced to additional system features that had not been used in Part 1, including features for sending feedback messages to groups during live monitoring.

This part served two purposes. For RQ1, it complemented the objective measures from Part 1 with participants' perceived ability to monitor concurrent discussions and their perceived workload during sustained use. Following the task, participants completed a post-task survey assessing their monitoring ability and workload. For RQ2, the continuous monitoring setting allowed us to examine how participants foraged for conversational evidence when they could not inspect every group or event immediately. We analyzed participants' screen-recorded interactions and conducted brief follow-up interviews to clarify the reasoning behind their decisions.

\section{Results}

We organize our findings around the two research questions. RQ1 asks, \textit{How does \sys{} affect users' ability to locate and inspect relevant conversational evidence during live sensemaking across concurrent discussions?} RQ2 asks, \textit{What evidence-foraging strategies do users adopt while making sense of multiple concurrently evolving conversations with \sys{}?}

\subsection{RQ1: Evidence-Foraging Performance}

\subsubsection{Objective Performance}

\begin{table}[h]
\centering
\caption{Monitoring performance from the user study across \sys{} (A), \sys{}-Content (B), and the transcript baseline (C). Best results per task are shown in bold. For condition A, * denotes \(p < 0.05\) and ** denotes \(p < 0.01\), with significance evaluated against condition B.}
\label{tab:perf_metrics}
\small
\setlength{\tabcolsep}{7pt}
\renewcommand{\arraystretch}{1.25}

\begin{tabular}{>{\raggedright\arraybackslash}p{4.2cm} | ccc | ccc | ccc}
\toprule
 & \multicolumn{3}{c|}{\textbf{Time (s)}} 
 & \multicolumn{3}{c|}{\textbf{Precision}} 
 & \multicolumn{3}{c}{\textbf{Recall}} \\

\textbf{Quiz Question} 
& \textbf{(A)} & \textbf{(B)} & \textbf{(C)} 
& \textbf{(A)} & \textbf{(B)} & \textbf{(C)} 
& \textbf{(A)} & \textbf{(B)} & \textbf{(C)} \\

\midrule

\rowcolor[HTML]{F2F2F2}
\multicolumn{10}{l}{\textit{Collaboration-Relevant Behaviors}} \\

Identify groups with conflicting ideas
& \textbf{19.78**} & 142.48 & 138.01
& \textbf{0.95} & 0.85 & 0.85
& \textbf{0.95} & 0.75 & 0.76 \\

Identify hesitant groups
& \textbf{18.46**} & 281.58 & 288.10
& \textbf{0.94*} & 0.75 & 0.76
& \textbf{0.96*} & 0.50 & 0.52 \\

\midrule

\rowcolor[HTML]{F2F2F2}
\multicolumn{10}{l}{\textit{Discussion Content}} \\

Concept coverage across groups
& 23.12 & \textbf{21.74} & 220.95
& 0.93 & 0.97 & \textbf{1.00}
& \textbf{0.80} & \textbf{0.80} & 0.56 \\

Most/least discussed ideas
& \textbf{25.79} & 26.34 & 204.31
& \textbf{1.00} & 0.80 & 0.40
& \textbf{1.00} & 0.80 & 0.40 \\

\bottomrule
\end{tabular}
\end{table}

We measured evidence-foraging performance using \textit{Time Spent}, \textit{Precision}, and \textit{Recall} (Table~\ref{tab:perf_metrics}). Overall, \sys{} (A) substantially improved the efficiency with which participants identified collaboration-relevant behaviors while also improving accuracy compared with conditions B and C.

The largest difference appeared when participants identified hesitant groups, which served as a representative case of monitoring collaboration-relevant acoustic cues. Participants using conditions B and C primarily relied on textual indicators such as filler words (e.g., ``uh''), which did not always appear and could instead reflect speakers searching for the right words rather than being hesitant to speak up. Reading through transcripts alone was also time-consuming, while listening to audio required participants to select a subset of groups without guidance on which moments were most relevant. With \sys{}, participants completed the task substantially faster (18.46~s vs. 281.58~s and 288.10~s) while achieving higher precision (0.94 vs. 0.75 and 0.76) and recall (0.96 vs. 0.50 and 0.52), as the system directly surfaced groups and moments associated with hesitation. For identifying groups with conflicting ideas, \sys{} similarly produced the highest precision (0.95) and recall (0.95), with an average completion time of 19.78~s compared with 142.48~s and 138.01~s in the other conditions. Text-based conditions (B and C) performed better on this task than on hesitation because participants could skim transcripts to locate potentially relevant moments without listening to audio, as this quiz represented collaboration-relevant lexical cues. However, doing so still required significantly more time than in condition A.

For discussion-content tasks, \sys{} and the ablated condition B performed similarly in both efficiency and accuracy. For example, identifying the most and least discussed ideas required 25.79~s with \sys{} and 26.34~s with the ablated condition. Precision and recall were also relatively close, with generally high values. These results are expected because both conditions provided the same discussion-content representations, with errors primarily occurring when participants inspected content from timestamps slightly before or after the designated segments, where neighboring or emerging content overlapped with the target discussion. Taken together, the results indicate that the primary benefit of the collaboration-relevant cue features was in locating and grounding conversational behaviors that were more difficult to infer directly from transcripts.

\subsubsection{Perceived Monitoring Ability and Workload}

\begin{figure*}
    \centering
    \includegraphics[width=1\linewidth]{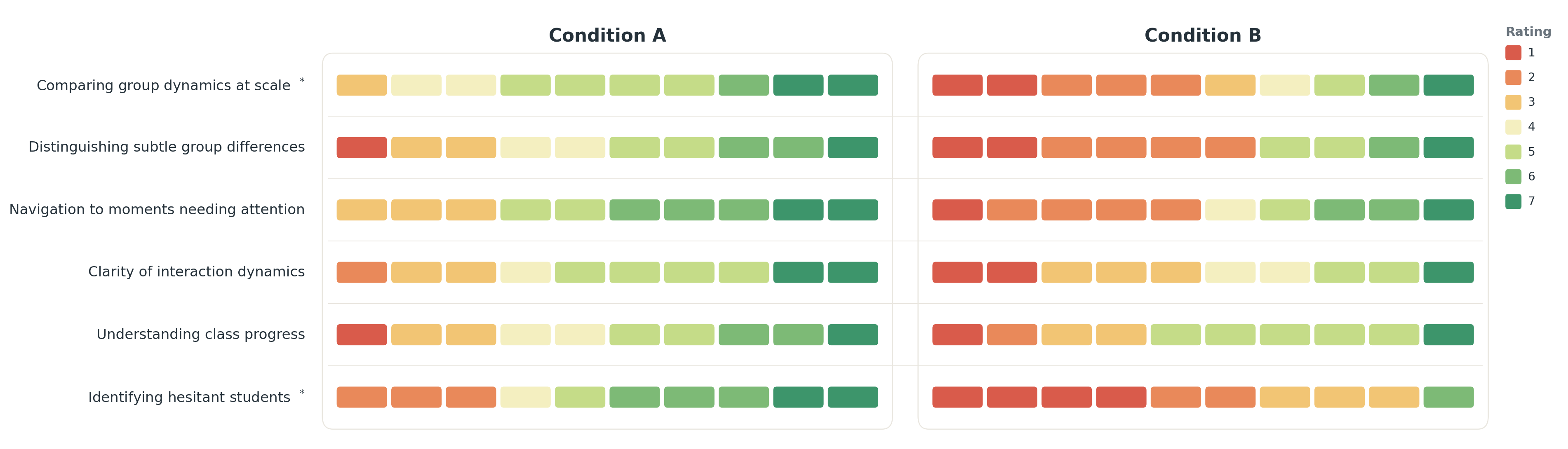}
    \caption{Participant ratings across six questions for \sys{} (Condition A; $N=10$) and the ablated \sys{} Discussion Content condition (Condition B; $N=10$). Each row shows individual responses on a 1--7 Likert scale, with higher values indicating more positive ratings. Asterisks indicate comparisons with $p<0.05$ using two-sided permutation Mann--Whitney U tests.}
    \label{fig:survey}
\end{figure*}

Between-condition differences were assessed using two-sided permutation Mann-Whitney U tests. Survey responses showed consistent patterns favoring \sys{} for tasks involving interactional evidence (Fig.~\ref{fig:survey}). Participants rated \sys{} higher in their ability to compare interaction patterns across groups at scale (A: 5.10 vs. B: 3.30; $U=76.0$, $p=0.0469$), distinguish subtle differences between groups with similar discussion patterns (A: 4.40 vs. B: 3.30), and navigate to specific moments or behaviors requiring attention (A: 5.10 vs. B: 3.70). Participants also reported that interactional information was more clearly represented with \sys{} (A: 4.60 vs. B: 3.60). P15 said the categories of cues helped them ``focus on opinions or disagreement,'' which was their priority when monitoring groups, while P7 highlighted annotations as ``very helpful'' for pinpointing what was happening at a particular moment. P19 also valued the visualization of cue amplitudes in the plot to convey their ``intensity'' through the patterns in which those cues appeared.

Participants similarly reported greater ability to identify hesitant students (reverse-coded; A: 4.70 vs. B: 2.30; $U=81.5$, $p=0.0130$), which served as a representative measure involving acoustic cues. Ratings related to discussion content understanding were more similar across conditions, such as participants' ability to identify groups that did not discuss the expected content (reverse-coded; A: 3.80 vs. B: 3.20). Participants using the full system also reported slightly higher confidence in their understanding of overall class progress (A: 4.40 vs. B: 4.10). Together, these results suggest that the additional interactional representations supported more detailed monitoring of conversational behaviors without reducing participants' broader understanding of class-wide progress.

Regarding workload, NASA-TLX responses did not exhibit statistically significant differences between conditions ($p>0.05$). The results were similar for mental demand (A: 4.60 vs. B: 5.00), physical demand (A: 3.00 vs. B: 2.60), effort (A: 4.50 vs. B: 4.90), temporal demand (A: 4.70 vs. B: 4.10), perceived performance (A: 3.70 vs. B: 3.90), and frustration (A: 3.00 vs. B: 3.50). While some dimensions were lower with \sys{}, others were slightly higher, which was expected because the full system presented participants with more interactional signals to monitor. P11 described seeing all groups at the same time within a limited period as ``challenging''. P7 mentioned experiencing high ``context switch load,'' as they frequently switched between groups and had to catch up with what the newly selected group was discussing. Nevertheless, based on the overall results, the additional collaboration-relevant cue representations supported CR monitoring without producing a significant increase in reported workload.

\subsection{RQ2: Evidence-Grounded Sensemaking Strategies in Live Conversations}

The sensemaking literature highlights several tradeoffs, such as those concerning how broadly people search, how deeply they process selected information, and where they allocate limited attention during sensemaking \cite{pirolli2005sensemaking}. We therefore conducted a theoretically informed analysis of the study recordings, using these tradeoffs as an analytic lens to examine the strategies that emerged as participants used \sys{} to forage conversational evidence. We focus on how these strategies unfolded when the underlying evidence was multimodal and continuously changing.

One pattern that became apparent from the recordings was that participants differed in what they sought to accomplish through sensemaking. Among the ten participants using Condition A, seven primarily sought to understand what groups were discussing and how their conversations developed, while three primarily focused on identifying opportunities to provide feedback. Five participants visibly shifted between these goals during the session. These changing goals shaped how they interacted with the system.

\subsubsection{Adjust Grounding Depth to the Sensemaking Goal}

Regarding the tradeoff between broadly exploring an information space and investing deeper attention in a smaller subset, all Condition A participants exhibited this broad-to-selective pattern. They scanned across the Group Monitoring Gallery, leaving many groups at the visual level while selectively investigating those that appeared relevant. \textit{Opinion}, \textit{Hesitation}, and \textit{Disagreement} were the three most frequently inspected collaboration-relevant cues.

With multimodal data, deeper processing involved evidence with different attentional costs, ranging from visual representations to transcripts and raw audio. Participants therefore did not consistently progress toward the richest available evidence. For example, P3 noticed sustained \textit{Uncertainty} in Group 2, hovered over the three most recent uncertainty cues to skim their transcripts, and then listened to the audio associated with all three moments. In contrast, P11 treated \textit{Interruption} cues as sufficient evidence for action, repeatedly sending feedback when new interruption cues appeared without reconstructing the discussion in greater depth. Other participants similarly acted on readily interpretable patterns such as silence or uneven speaker participation. These behaviors suggest that participants sought sufficient evidence for their current goal.

\subsubsection{Trace Cues Back Through Conversation History}

Sensemaking can be described as an opportunistic mixture of bottom-up and top-down processing, where evidence can generate an interpretation and an interpretation or question can initiate a new search for evidence.

Conversational time served as an important dimension for processing interpretations. While bottom-up processing may have been more common with \sys{}, top-down processing was also evident. Eight out of ten participants inspected earlier portions of a group's conversation during the session. For example, P13 noticed a \textit{Disagreement} cue in Group 1 and then inspected the \textit{Opinion} cue immediately preceding it to understand the position that had prompted the disagreement.

\subsubsection{Defer Grounding to Keep Up with Live Conversations}

The costs of shifting attention and the limits on how much evidence can be maintained simultaneously are important constraints on sensemaking. These constraints were amplified in our setting because attending to one group did not pause the others. Although \sys{} allowed participants to leave audio playing while visually browsing other groups, they still had to make sense of multiple streams of continuously evolving information.

To deal with evolving data, participants therefore sometimes deferred grounding when their attention was occupied elsewhere, relying on the persistent timeline to return to potentially relevant evidence later. P11, for instance, initially noticed a \textit{Hesitation} cue in Group 3, hovered over it to skim the transcript, and continued scrolling through other groups. After returning to Group 3 and seeing that additional cues had appeared, P11 went back to the original cue, listened to its audio, and then listened to the newly emerged cue to understand how the discussion had developed.

\section{Discussion}

\subsection{Toward AI Systems for Understanding Tacit Collaboration Across Modalities}

As AI increasingly moves toward models that represent richer aspects of the world, an important frontier is how such models represent the humans acting within that world and, particularly, how people collaborate. While systems have begun characterizing different aspects of collaboration, such as group discussions \cite{CPVis,groupCommAnalysis}, many of the signals people use to collaborate with one another can be tacit rather than explicit \cite{tacitCoor}. Consider an in-person hardware assembly task, where one collaborator might subtly reposition or tilt a component and another immediately understands that they should hold it, align another part, or prepare for the next action. The movement itself may be easy to sense, but understanding what it means for the collaboration might not be as straightforward. As AI gains access to increasingly rich multimodal traces of human activity, HCI has an important role in identifying which of these human-recognized distinctions are meaningful and useful for enhanced collaboration.

\sys{} takes a step in this direction. In this work, we sought to characterize \textit{Hesitation} and \textit{Uncertainty}, whose manifestations are less directly interpretable and required us to translate prior knowledge into annotation rubrics and refine those rubrics through human interpretation. \sys{} presents these moments as collaboration-relevant cues that instructors can interpret in context, while preserving access to the underlying raw audio evidence to support further sensemaking through human expertise. Looking forward, multimodal AI may enable systems to represent other tacit collaboration dynamics across speech, gesture, gaze, and many other modalities. We see an important role for HCI in determining how such dynamics should be operationalized and represented while keeping their interpretation contextual and open to human judgment.

\subsection{Design Implications}

Our findings highlight a tension specific to evidence-grounded live sensemaking: users need enough temporal context to interpret conversational evidence, but inspecting that evidence consumes attention while other conversations continue to evolve. The temporal signal density plot in \sys{} embodies one way to balance these demands by making the temporal distribution of collaboration-relevant cues visible at a glance. Rather than presenting detected cues only as isolated events, temporal representations can help users distinguish one-off occurrences from recurring or sustained patterns before deciding whether deeper inspection is warranted. Prior systems similarly demonstrate the value of temporal representations for understanding evolving collaborative behavior \cite{TalkTraces,CPVis} and compact visual summaries for comparing multiple concurrent groups \cite{Groupnamics,MobileGroupVis}. We therefore suggest that evidence-grounded live sensemaking systems use compact temporal representations that preserve signal patterns while making intervals that may warrant attention readily apparent. Importantly, these representations should function as entry points into evidence rather than final interpretations, allowing users to decide when further grounding is necessary.

Second, intermediate representations positioned between transcripts and raw audio can reduce the cost of accessing conversational evidence. Transcription can remove information carried by the original audio \cite{WavRAG}, while speech-based systems have also shown that raw transcripts can be difficult to interpret because spoken language contains disfluencies, repetitions, and other characteristics that are not easily represented as structured information \cite{Orality}. At the same time, providing more evidence does not necessarily make sensemaking easier: prior work on evidence provenance and AI transparency highlights the cognitive cost of inspecting detailed supporting information and the value of providing detail when it becomes relevant \cite{PaperTrail,GoldilocksTransparency}. Intermediate representations can provide this bridge by supporting lightweight inspection before users commit additional attention to source audio. For example, hover-based access to transcripts and associated conversational cues can help users determine whether a moment warrants further investigation before listening. We therefore suggest designing evidence access progressively, allowing users to move from compact AI-generated representations, to contextual textual evidence, and finally to the original source when additional grounding is needed.

\section{Limitations \& Future Work}

This work has several limitations. First, our main evaluation was conducted in a controlled laboratory setting and does not capture how \sys{} would affect classroom practice or student learning over sustained use. Although we deployed the system in live classes, those deployments primarily served as an end-to-end technical evaluation of the complete pipeline. Future work should evaluate \sys{} longitudinally across a semester to understand how it influences instructor monitoring and broader teaching practices.

Second, the current system captures a set of interactional and paralinguistic cues. Our implementation focuses on hesitation and uncertainty as acoustic signals, alongside lexical and speaker-timing cues. Future work should investigate additional paralinguistic and interactional cues, such as prosodic characteristics and laughter, which prior work has linked to social interaction and group effectiveness \cite{vinciarelli2009social, woolley2010evidence}.

Finally, although the audio data in our study was anonymized, collecting and processing classroom speech raises important privacy and data-access concerns \cite{aiTeacherCollabInClass}. Future work should explore privacy-preserving approaches such as on-device processing, together with mechanisms that provide instructors and students greater transparency and control over how speech data is captured and used.

\section{Conclusion}
We presented \sys{}, an evidence-grounded Human-AI sensemaking system for monitoring multiple concurrent classroom CR discussions, developed through our formative study. \sys{} introduces a temporal signal density visualization that compresses detected collaboration-relevant cues into a compact temporal representation, allowing instructors to quickly distinguish patterns while preserving when those events occurred within the evolving conversation. Our mixed-design study with 20 participants showed that \sys{} improved the efficiency and accuracy of grounding interactional interpretations, while participants used strategies such as deferred grounding, evidence revisitation, and progressive grounding to navigate evidence across evolving conversations. More broadly, our work studies how AI-generated abstractions can support not only the summarization of live conversational data, but also efficient foraging for evidence and navigation back to underlying sources, including conversational audio, when richer context is needed for human sensemaking.

\section{Generative AI Usage}
Generative AI tools were used to support prototype implementation, data analysis, and manuscript preparation. Claude and OpenAI Codex assisted with code development for the research prototype. ChatGPT was used to assist in drafting Python scripts for processing survey data and computing the statistical analyses reported in this paper. All AI-assisted code and analysis procedures were reviewed, modified as necessary, and executed by the authors. ChatGPT was also used to support the writing of this manuscript and the partial creation of its teaser figure.

\begin{acks}
This work is funded by TBD. We would like to thank TBD for their help with data collection and providing feedback. We would like to thank TBD with the help with the demo. We would like to thank TBD for their feedback on the project. 
\end{acks}

\bibliographystyle{ACM-Reference-Format}
\bibliography{sample-base}

\appendix
\section{Dataset Collection Details}
\label{appendix:dataCollection}

\begin{figure}[H]
    \centering
    \includegraphics[width=\columnwidth]{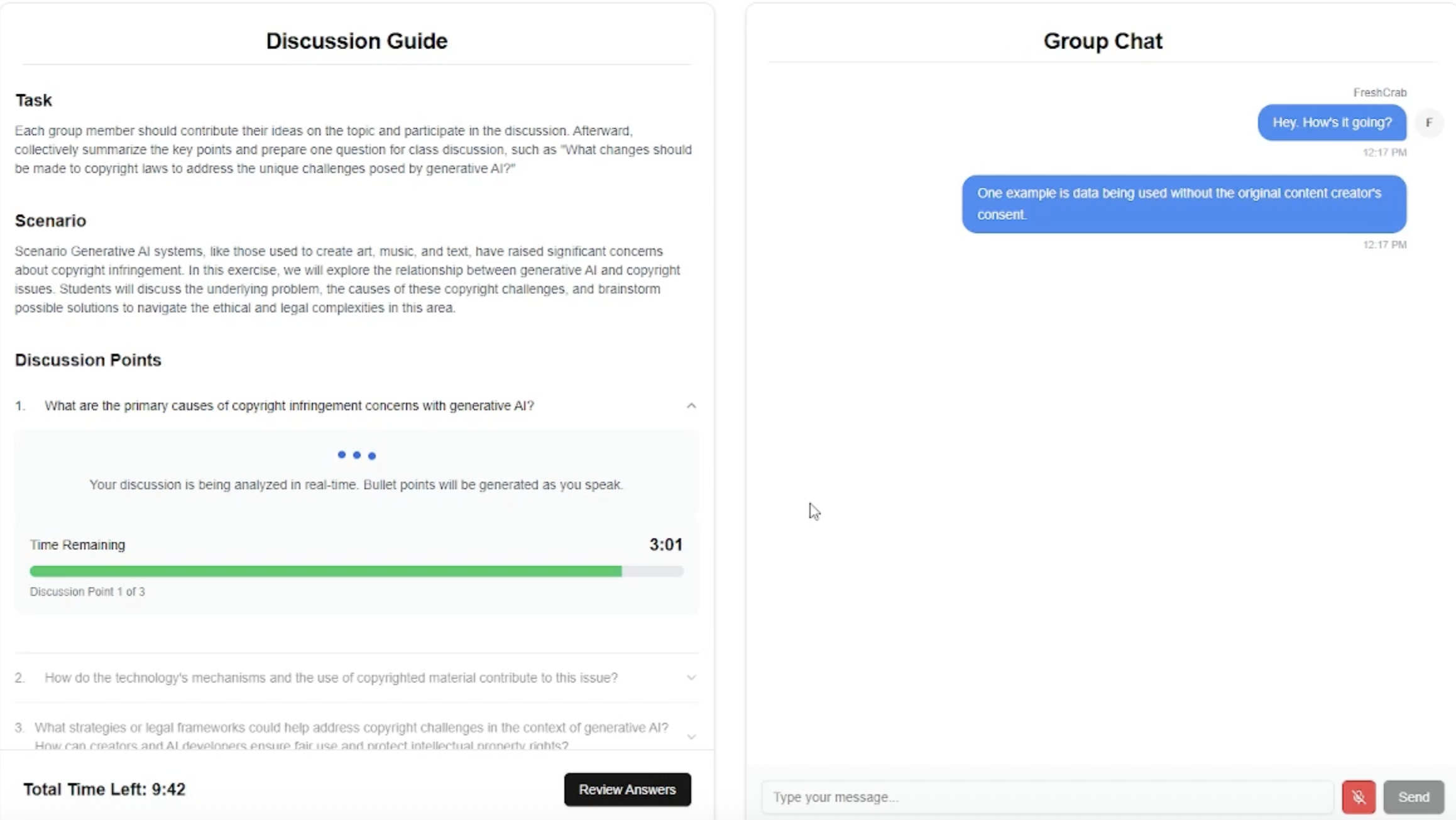}
    \caption{Student-facing interface showing the discussion task description, guiding questions, and remaining time for that discussion point on the left. On the right, students can view live transcripts of group members.}
    \label{fig:studentScreenshot}
\end{figure}

The data collection process began after we observed several classes taught by E1 and E2. We developed a data collection system and worked with instructors E1 and E2 to deploy it in their classes. We deployed the system in one of E1's classes with 85 students and in two of E2's classes with 35 and 34 students, respectively. Before each class, we worked with E1 and E2 to create discussion tasks that aligned with their course agendas. We informed the instructors that the data collection was approved by our institution's IRB.

During each class, two members of our research team attended, including the first author. The instructor introduced the research team, and the first author explained the data collection process, shared the IRB approval letter with the class, and clarified that participation was voluntary, would have no consequences for students' grades, and would not involve collecting personally identifying information. Students were also informed that their voices would be anonymized. Students who did not wish to participate could continue the activity by discussing in Zoom as usual. Consent was collected through the system's landing page. Students who provided informed consent could then enter their assigned breakout room number to participate (Fig.~\ref{fig:studentScreenshot}). During the group activity, the instructor and the research team remained in the main Zoom room, where we gathered feedback on the initial version of the instructor-facing interface.

\section{Participant Condition Assignment}
\label{appendix:conditionAssignment}

Table~\ref{tab:participantConditions} summarizes the condition assignment and presentation order for each participant. 

\begin{table*}[t]
\centering
\caption{Participant condition assignments across the two study parts. Condition A denotes the full \sys{} system, Condition B denotes the ablated \sys{} Discussion Content condition, and Condition C denotes the transcript baseline.}
\label{tab:participantConditions}
\small
\setlength{\tabcolsep}{8pt}
\renewcommand{\arraystretch}{1.15}
\begin{tabular}{lcc}
\toprule
\textbf{Participant} & \textbf{Part 1 Order} & \textbf{Part 2 Condition} \\
\midrule
P1  & A $\rightarrow$ C & A \\
P2  & B $\rightarrow$ C & B \\
P3  & C $\rightarrow$ A & A \\
P4  & C $\rightarrow$ B & B \\
P5  & C $\rightarrow$ A & A \\
P6  & C $\rightarrow$ B & B \\
P7  & A $\rightarrow$ C & A \\
P8  & B $\rightarrow$ C & B \\
P9  & A $\rightarrow$ C & A \\
P10 & B $\rightarrow$ C & B \\
P11 & C $\rightarrow$ A & A \\
P12 & C $\rightarrow$ B & B \\
P13 & C $\rightarrow$ A & A \\
P14 & A $\rightarrow$ C & A \\
P15 & C $\rightarrow$ A & A \\
P16 & C $\rightarrow$ B & B \\
P17 & B $\rightarrow$ C & B \\
P18 & C $\rightarrow$ B & B \\
P19 & A $\rightarrow$ C & A \\
P20 & B $\rightarrow$ C & B \\
\bottomrule
\end{tabular}
\end{table*}

\section{Lexical Cue Classification Prompts}
\label{appendix:lexicalPrompts}

\subsection{Opinion Cue}

For each incoming utterance, the model receives the preceding conversational context and determines whether the current utterance introduces an \textit{Opinion Cue}.

\begin{verbatim}
You are analyzing a collaborative group discussion.

Task:
Determine whether the CURRENT UTTERANCE contains an
Opinion Cue.

Definition:
An Opinion Cue occurs when a speaker introduces a claim,
factual or reference information, or another substantive
idea that is new relative to ideas expressed previously
in the conversation.

Use the preceding conversation to determine whether the
idea introduced in the current utterance is new.

Few-shot examples:
[Discussion Tracker examples used in the implementation]

Preceding conversation:
{CONVERSATION_CONTEXT}

Current utterance:
Speaker: {SPEAKER_ID}
Transcript: {TRANSCRIBED_UTTERANCE}
\end{verbatim}

\subsection{Disagreement Cue}

\sys{} first accumulates consecutive utterances from the same speaker into a speaker turn. The resulting turn and its preceding conversational context are then provided to the classifier.

\begin{verbatim}
You are analyzing a collaborative group discussion.

Task:
Determine whether the CURRENT SPEAKER TURN contains a
Disagreement Cue.

Definition:
A Disagreement Cue occurs when a speaker questions,
challenges, or expresses disagreement with an idea
introduced in a prior speaker turn.

A speaker turn is a continuous contribution from one
speaker before another speaker takes the floor. A current
turn may disagree with multiple preceding turns, but the
same prior turn should not be counted as a disagreement
more than once.

Few-shot examples:
[Discussion Tracker examples used in the implementation]

Preceding conversation:
{CONVERSATION_CONTEXT}

Current speaker turn:
Speaker: {SPEAKER_ID}
Transcript: {SPEAKER_TURN}
\end{verbatim}

\subsection{Coordination Cue}

For \textit{Coordination}, each incoming utterance is classified independently.

\begin{verbatim}
You are analyzing a collaborative group discussion.

Task:
Determine whether the CURRENT UTTERANCE contains a
Coordination Cue.

Definition:
A Coordination Cue occurs when a speaker uses
meta-discourse to manage or direct the group discussion.

Examples of coordination include:
- introducing a discussion question,
- inviting another participant to contribute, or
- proposing a transition between discussion points.

Few-shot examples:
[Researcher-labeled examples used in the implementation]

Preceding conversation:
{CONVERSATION_CONTEXT}

Current utterance:
Speaker: {SPEAKER_ID}
Transcript: {TRANSCRIBED_UTTERANCE}
\end{verbatim}

\end{document}